\documentclass[a4paper]{spie}  

\usepackage{amsmath,amsfonts,amssymb}
\usepackage{graphicx}
\usepackage[colorlinks=true, allcolors=blue]{hyperref}
\usepackage{xcolor}
\usepackage{tikz}
\usepackage{multirow,array}

\DeclareRobustCommand{\swatch}[1]{\tikz[baseline=-0.7ex] \node[fill=#1,shape=rectangle,draw=#1,thick,minimum width=2.0mm,rounded corners=2pt](){};}

\title{Segmentation of Retinal Low-Cost Optical Coherence Tomography Images using Deep Learning}

\author[a,b]{Timo Kepp}
\author[c]{Helge Sudkamp}
\author[e]{Claus von der Burchard}
\author[e]{Hendrik Schenke}
\author[c]{Peter Koch}
\author[c,d]{Gereon H\"uttmann}
\author[e]{Johann Roider}
\author[a]{Mattias P. Heinrich}
\author[a]{Heinz Handels}

\affil[a]{Institute of Medical Informatics, University of L\"ubeck, Germany}
\affil[b]{Graduate School for Computing in Medicine and Life Sciences, University of L\"ubeck}
\affil[c]{Medical Laser Center L\"ubeck, Germany}
\affil[d]{Institute of Biomedical Optics, University of L\"ubeck, Germany}
\affil[e]{Department of Ophthalmology, University of Kiel, Germany}

\authorinfo{Send correspondences to Timo Kepp: kepp@imi.uni-luebeck.de}

\begin{document} 
\maketitle

\begin{abstract}
The treatment of age-related macular degeneration (AMD) requires continuous eye exams using optical coherence tomography (OCT). The need for treatment is determined by the presence or change of disease-specific OCT-based biomarkers. Therefore, the monitoring frequency has a significant influence on the success of AMD therapy. However, the monitoring frequency of current treatment schemes is not individually adapted to the patient and therefore often insufficient. While a higher monitoring frequency would have a positive effect on the success of treatment, in practice it can only be achieved with a home monitoring solution. One of the key requirements of a home monitoring OCT system is a computer-aided diagnosis to automatically detect and quantify pathological changes using specific OCT-based biomarkers. In this paper, for the first time, retinal scans of a novel self-examination low-cost full-field OCT (SELF-OCT) are segmented using a deep learning-based approach. A convolutional neural network (CNN) is utilized to segment the total retina as well as pigment epithelial detachments (PED). It is shown that the CNN-based approach can segment the retina with high accuracy, whereas the segmentation of the PED proves to be challenging. In addition, a convolutional denoising autoencoder (CDAE) refines the CNN prediction, which has previously learned retinal shape information. It is shown that the CDAE refinement can correct segmentation errors caused by artifacts in the OCT image.
\end{abstract}


\section{Purpose}
\label{sec:purpose}
Age-related macular degeneration (AMD) is the leading cause of blindness in the western world \cite{Bourne2013}. AMD causes damages to the macula, the central part of the retina that enables the sharp vision and thus is crucial for reading and face recognition. Optical coherence tomography (OCT) is the standard imaging modality for the assessment of AMD. It is used to create volumes composed of cross-sectional 2D images (B-scans) of the retina, allowing detailed analysis of retinal layers as well as the detection of pathological changes. Morphological OCT-based biomarkers such as intraretinal cysts (IRC), subretinal fluid (SRF) or pigment epithelial detachment (PED) are treatment indicators of AMD therapy \cite{Pauleikhoff2013}. In addition, it has been shown that retinal volume change may also be a reliable biomarker for the activity of AMD \cite{Burchard2018}.

The monitoring frequency plays an important role in the success of AMD treatment. The examination intervals of current treatment schemes are often too large and are not adapted to the patient's individual recurrence pattern. Therefore, it cannot be guaranteed that pathological changes will be diagnosed at an early stage. In order to improve the therapy, the monitoring frequency should be increased significantly, e.g. to a weekly or even daily interval. However, shorter examination intervals in the clinic would result in a high burden for the patient and would at the same time lead to high personnel costs. Therefore, in practice, a higher monitoring frequency can only be achieved by a home monitoring solution, i.e. by a low-cost OCT system at the patient's home for self-examination. A novel concept of a self-examination low-cost full-field OCT (SELF-OCT) has recently been developed as a prototype \cite{Sudkamp2016, Sudkamp2018} and is currently being tested in clinical trials. One essential requirement of the SELF-OCT as a home solution for AMD monitoring is an automated computer-aided evaluation of the image data in order to be able to track the progress of the disease. With regard to the expected high amount of image data, any manual assessment is not feasible.

In recent years, algorithms for the segmentation of OCT-based biomarkers have been described in several works \cite{Wintergerst2017}. Among the state-of-the-art methods, the number of deep-learning-based techniques such as convolutional neural networks (CNNs) is constantly increasing. Deep learning methods for segmentation of the total retinal segmentation \cite{Venhuizen2017}, retinal fluids like IRC, SRF \cite{Breger2017} or PED \cite{Xu2017} have been proposed. Furthermore, the RETOUCH challenge was recently organized to measure the performance of state-of-the-art methods for the detection and segmentation of retinal fluids in OCT \cite{Bogunovic2019}. In this work, for the first time, the total retina volume as well as PED is segmented in SELF-OCT image data using a deep learning-based approach. The segmentation approach builds on our preliminary work \cite{Kepp2019a, Kepp2019b} and consists of a CNN that segments the total retina as well as PEDs in three-dimensional scans of the SELF-OCT and is based on the popular U-Net architecture \cite{Ronneberger2015}. In comparison to commercially available clinical OCT systems, the scans of the SELF-OCT show a lower quality with a reduced signal-to-noise ratio (SNR). Furthermore, motion artifacts can impair image interpretation and lead to segmentation errors. To tackle this problem, the integration of anatomical prior knowledge into the segmentation process was investigated in previous works \cite{Ravishankar2017, Oktay2017, Larrazabal2019}. In their work \cite{Ravishankar2017}, Ravishankar et al. used a convolutional denoising autoencoder (CDAE) to learn anatomical shapes so that corrupt segmentation can be corrected. For this purpose, the pre-trained CDAE was cascaded to the segmentation network. They evaluated their approach by segmenting kidneys in longitudinal ultrasound images. In another work, Larrazabal et al. \cite{Larrazabal2019} also used an autoencoder for the post-processing of lung segmentation in X-rays. To correct corrupt predictions of the U-Net, we use a CDAE as a refinement step so that the segmentation approach becomes robust against image artifacts and reduced SNR.

\section{Methods}
\label{sec:methods}
\subsection{Self-Examination Low-Cost Full-Field OCT of the Retina}
\label{sec:lcoct}
\begin{figure}[t]
	\centering
	\begin{tabular}{cc}
		\includegraphics[width=0.48\textwidth]{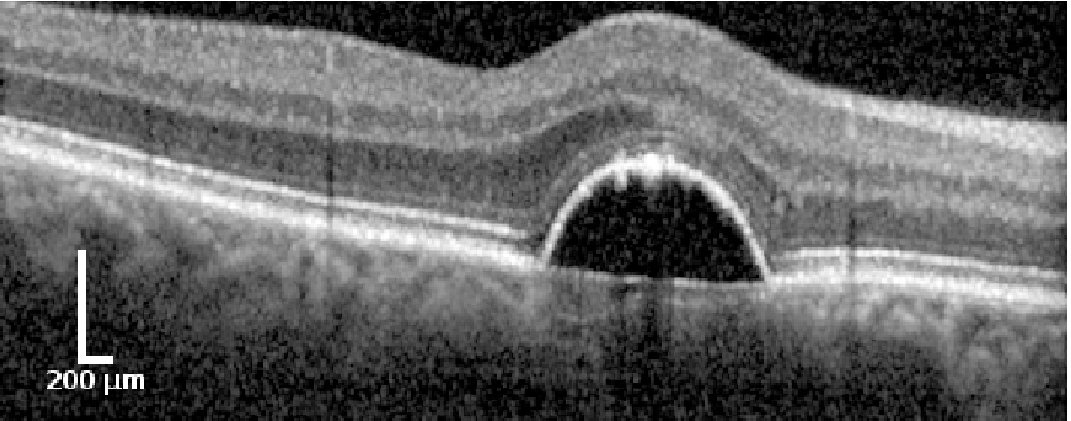} & \includegraphics[width=0.465\textwidth]{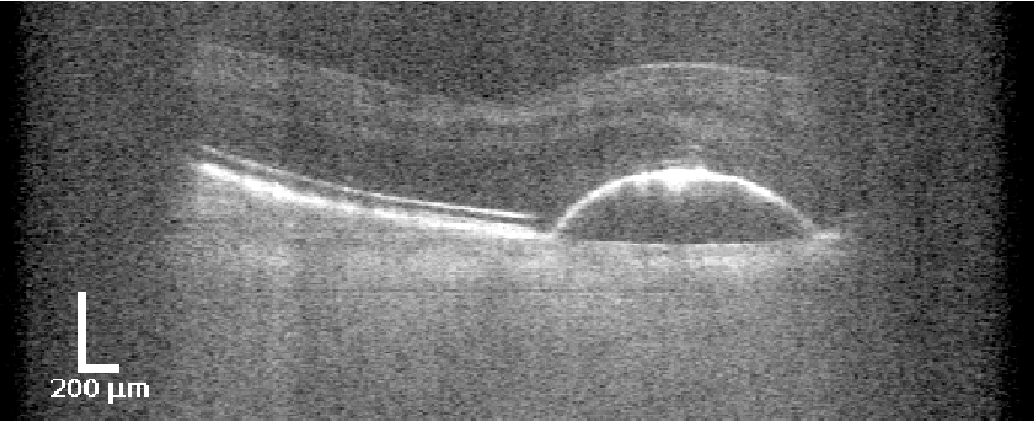}\\
    \end{tabular}
    \vspace{0.5em}
	\caption{Comparison of two retinal scans of the same AMD patient with PED between Spectralis OCT from Heidelberg Engineering (left) and SELF-OCT (right).\label{fig:spectralisandlowcost}}	
\end{figure}

The SELF-OCT is a low-cost OCT system that allows the patient to independently examine the disease progression of AMD at home without the presence of a physician. In contrast to commercially available clinical OCT systems, the SELF-OCT system sequentially acquires transversal \textit{en face} images at different depths instead of cross-sectional images in axial direction \cite{Sudkamp2016}. The imaging geometry results in a dense lateral sampling (compared to a dense lateral sampling in the first axis and a sparse sampling in the second axis, which is usually used in OCT imaging). With modern CMOS cameras and a simple broadband light source, cost-effective and compact OCT systems can be built that are readily accessible and easy to operate for the patient. However, it must also be taken into account that scans from the SELF-OCT differ from those from clinic OCTs in terms of artifacts and SNR, which leads to a different image impression (Fig. \ref{fig:spectralisandlowcost}). Furthermore, the field of view of the SELF-OCT is smaller compared to clinical OCT systems as a results of the fast acquisition time ($\sim 0.9$ seconds per volume). However, the field of view is sufficient for clinical diagnosis of common AMD biomarkers \cite{Burchard2017}.

\subsection{Dataset}
\label{sec:dataset}
In a clinical pilot study image data of a group of 51 patients were acquired by the SELF-OCT system. In addition to the study eye, the second eye of each patient was also scanned, which doubled the number of image data. In addition to AMD patients, a small number of patients with other retinal diseases such as diabetic macular edema (DME), retinal vein occlusion (RVO) or central serous retinopathy (CSR) were included in the pilot study. Moreover, the partner eye was mostly also diseased. All OCT volumes have an axial resolution of 9.1\,$\mu$m and a transversal resolution of 6.4\,$\mu$m and 12.8\,$\mu$m, respectively. The volume size is 250$\times$496$\times$140\,px with a field of view size of 1.4$\times$4.8$\times$1.5\,mm. The SELF-OCT acquires multiple image volumes during each scanning process so that on average 10 repetitive scans per eye were generated. To improve the SNR the OCT volumes were processed with a moving average filter of size five. All OCT scans were subsequently evaluated by an independent expert with regard to their quality. Based on this evaluation high-quality scans were selected by a medical expert and manually annotated. In order to keep the annotation work in a feasible scope, the image size of each volume was reduced to one fifth in one lateral direction so that a total of only 28 B-scans had to be segmented per volume. All image voxels were assigned to one of the following three classes: Retina (\swatch{red}), PED (\swatch{green}) and background (no color). The retina is defined as the area between the inner limiting membrane (ILM) and the Bruch's membrane (BM). The PED is delimited above by the retinal pigment epithelium (RPE) and below by the BM. An example of a manual segmentation is shown in Fig. \ref{fig:qualitativeresult}. In total, the data set consists of 711 image volumes with corresponding ground truth annotation from 44 different patients. Image data with insufficient quality were excluded. The retinal diseases of the patient's eyes are composed as follows: 53~AMD, 7~DME, 3~RVO, 2~CSR, and 7~healthy.
\begin{figure}[h]
    \centering
	\includegraphics[width=\textwidth]{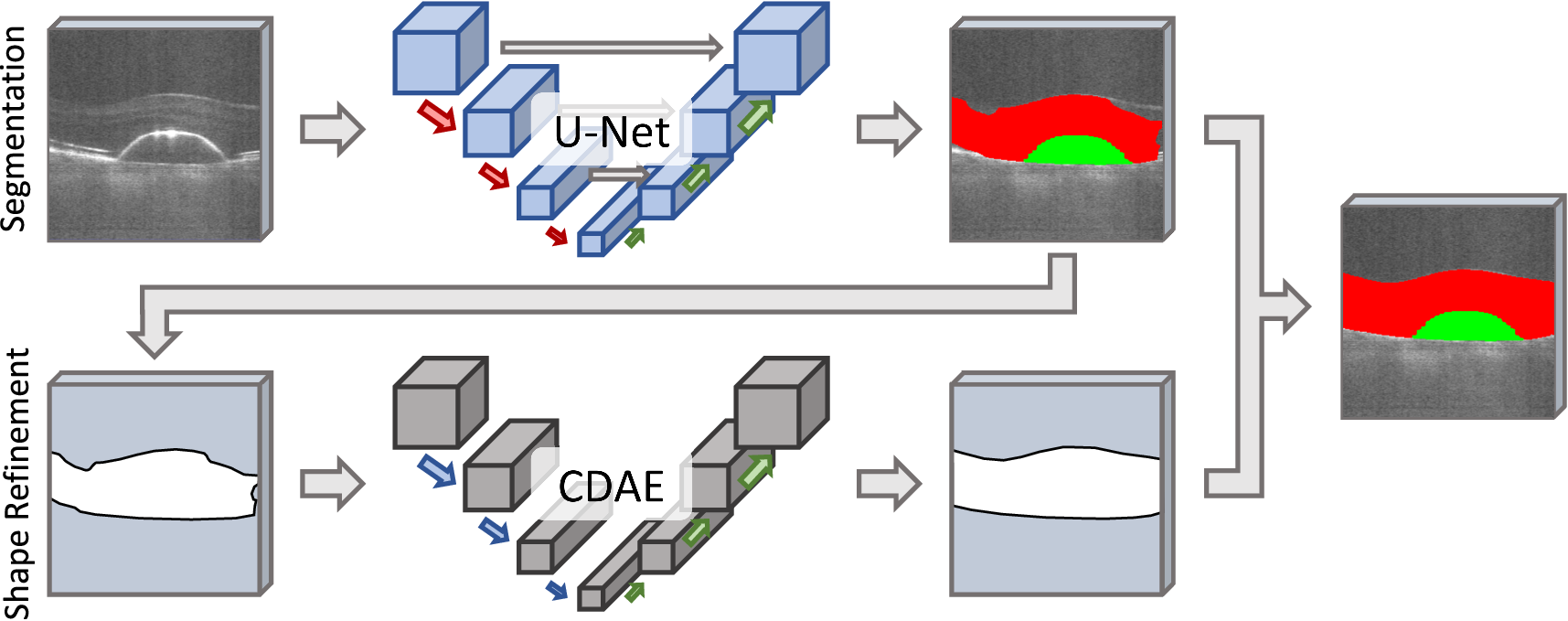}
	\vspace{0.5em}
	\caption{Schema of the segmentation pipeline. First row: multi-class segmentation of the OCT volume with the U-Net and generation of the retinal shape via binarization. Second row: Shape refinement by the CDAE and fusion with the U-Net segmentation. \label{fig:scheme}}	
\end{figure}

\subsection{Deep Learning-based Segmentation of Retinal SELF-OCT}
\label{sec:deeplearningbasedsegmentationofretinallcoct}
Quantitative evaluation of OCT-biomarkers is crucial for a home monitoring solution in AMD therapy. The segmentation of these biomarkers is an essential basis for this. In this paper, a segmentation framework was developed that automatically segments both the retina and the PED as biomarkers in SELF-OCT image data. Thus, the segmentation task was to classify each voxel of a SELF-OCT input scan into one of three classes: retina (\swatch{red}), PED (\swatch{green}) and background. An overview of the proposed segmentation framework is given in Fig. \ref{fig:scheme}. The framework consists of a segmentation and a refinement part. In a first step, the retina, as well as the PED, were segmented in the input scan using a 3D U-Net architecture \cite{Cicek2016}. One of the main challenges was the comparatively low image quality of the SELF-OCT compared to standard clinical OCT (Fig. \ref{fig:spectralisandlowcost}) as well as the occurrence of motion artifacts that can significantly influence the interpretability and lead to reduced segmentation accuracy. Therefore, in a second step, a shape refinement was conducted to correct segmentation errors that occurred due to artifacts. This was done by first binarizing the output of the U-Net and then passing it as input to the CDAE. By training the CDAE with binarized ground truth segmentations with added artificially generated errors, it learned to correct corrupt retinal shapes. To achieve this the non-corrupt ground truth version of the respective input image was used as a training target. In a final step, the corrected retinal shape was fused with the U-Net prediction. In the following, both the U-Net segmentation and the CDAE refinement will be described in more detail.

\subsubsection{3D Segmentation of the total retina and PED using the U-Net architecture}
\label{sec:3dunet}
The CNN used in this work is based on the U-Net, a popular CNN architecture used in biomedical image analysis, where it shows high accuracy in the field of segmentation \cite{Ronneberger2015}. This fully convolutional network consists of a contracting encoder that captures large contextual information on the one hand, and an expanding decoder part, which enables dense class predictions on the other hand. Due to the special acquisition technique described in Sec. \ref{sec:lcoct}, the SELF-OCT generates densely sampled volumes making the 3D U-Net architecture \cite{Cicek2016} a natural choice for OCT segmentation. It receives a complete OCT volume instead of single 2D B-scans as input and outputs a voxel-wise classification.

\begin{figure}[t]
    \centering
	\includegraphics[width=\textwidth]{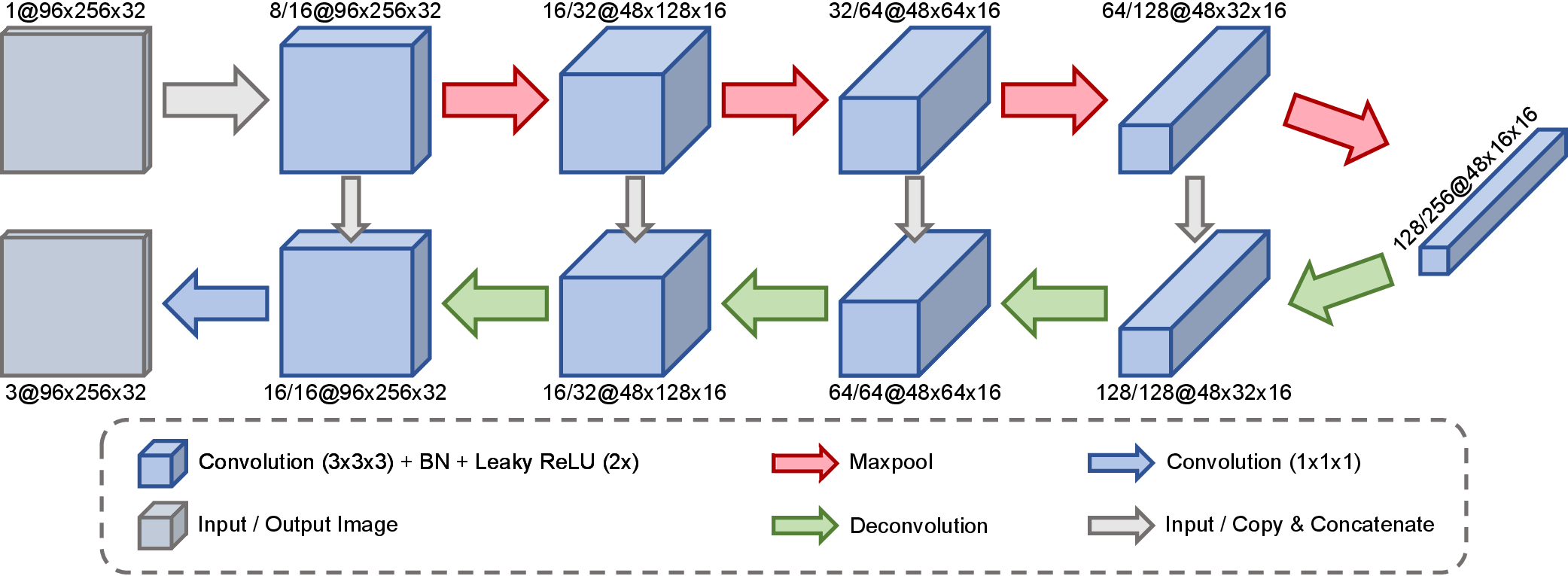}
	\vspace{0.3em}
	\caption{Architecture of the proposed 3D U-Net. The notation C$_1$/C$_2$@W$\times$H$\times$D at each blocks describes the number of output channels of the first or second convolution (C$_{1/2}$) for a given image resolution (width$\times$height$\times$depth). \label{fig:unet}}	
\end{figure}

The architecture of the 3D U-Net is shown in Fig. \ref{fig:unet}. Each scale level of the encoder consists of two 3$\times$3$\times$3 convolutions followed by a leaky rectified linear unit (ReLU) as an activation function. A 2$\times$2$\times$2 max pooling with a stride of two halves the resolution in the axial direction. In order not to lose detailed image information a stride of one is used for the two lateral dimensions except for the first scale level. The amount of feature channels increases gradually in the encoder path so that a number of 256 feature channels is reached at the lowest scale level. In the following decoder path, the image resolution is successively increased again using 2$\times$2$\times$2 deconvolutions, so that they correspond to those of the encoder path. Skip connections link the multiple scale levels of the encoder and decoder to ensure spatial consistency and increase gradient flow during training. In contrast to the encoder, only one 3$\times$3$\times$3 convolution is used per resolution level in the decoder path. The last layer consists of a 1$\times$1$\times$1 convolution which reduces the number of feature channels to the number of classes. In addition, batch normalization is applied after each 3$\times$3$\times$3 convolution to achieve faster convergence during training.

For the training of the 3D U-Net Adam optimization was used whose exponential decay rates $\beta_1$ and $\beta_2$ were set to $0.9$ and $0.999$, respectively. An initial learning rate of $10^{-3}$ was conducted which was continuously adjusted by an additional exponential scheduler with a decay rate of $\gamma=0.99$. The generalized Dice \cite{Sudre2017} was used as loss function, which quantifies the overlap between predicted and ground truth segmentation. To address the high class imbalance each class was weighted by its inverse class frequency during loss computation. The CNN was trained for 500 epochs because after this time the training loss converged. Additionally, due to the small amount of training data, online data augmentation was performed to avoid overfitting. Therefore, horizontal flips, similarity transforms and non-linear intensity shifts were randomly applied to the input volumes. 

\subsubsection{Shape-based Refinement Using a Convolutional Denoising Autoencoder}
\label{sec:shapebasedrefinementusingacae}
\begin{table}[b]
    \caption{Architecture of the convolutional denoising autoencoder. The configurations are defined as follows. k: kernel form, s: stride, f: upscaling factor, \#: number of output channels. \label{tab:cdae_architecture}}
    \vspace{0.5em}
    \begin{minipage}[t]{.45\textwidth}
        \centering
        \begin{tabular}{cclc}
            \multicolumn{4}{c}{Encoder} \\
            \hline
            & Layer & \multicolumn{1}{c}{Configuration} & Output shape \\
            \hline\
            
            \rotatebox[origin=c]{90}{$S_1$} & Conv & k:(3,3,3), s:(1,1,1), \#:8 & 96$\times$256$\times$32 \\
            \hline
            \multirow{2}{*}{\rotatebox[origin=c]{90}{$S_2$}} & Conv & k:(3,3,3), s:(1,2,1), \#:16 & 96$\times$128$\times$32 \\
                                                             & Conv & k:(3,3,3), s:(1,1,1), \#:32 & 96$\times$128$\times$32 \\
            \hline
            
            \multirow{2}{*}{\rotatebox[origin=c]{90}{$S_3$}} & Conv & k:(3,3,3), s:(1,2,1), \#:32 & 96$\times$64$\times$32 \\
                                                             & Conv & k:(3,3,3), s:(1,1,1), \#:32 & 96$\times$64$\times$32 \\
            \hline
             
            \multirow{2}{*}{\rotatebox[origin=c]{90}{$S_4$}} & Conv & k:(3,3,3), s:(2,2,2), \#:32 & 48$\times$32$\times$16 \\
                                                             & Conv & k:(3,3,3), s:(1,1,1), \#:64 & 48$\times$32$\times$16 \\
            \hline

            \multirow{2}{*}{\rotatebox[origin=c]{90}{$S_5$}} & Conv & k:(3,3,3), s:(2,2,2), \#:64 & 24$\times$16$\times$8 \\
                                                             & Conv & k:(3,3,3), s:(1,1,1), \#:64 & 24$\times$16$\times$8 \\
            \hline
            
            \multirow{2}{*}{\rotatebox[origin=c]{90}{$S_6$}} & Conv & k:(3,3,3), s:(2,2,2), \#:128 & 12$\times$8$\times$4 \\
                                                             & Conv & k:(3,3,3), s:(1,1,1), \#:128 & 12$\times$8$\times$4 \\
                                                             & Conv & k:(3,3,3), s:(1,2,1), \#:8 & 12$\times$8$\times$4 \\
            \hline
        \end{tabular}
    \end{minipage}
    \hfill
    \begin{minipage}[b]{.45\textwidth}
        \centering
        \begin{tabular}{cclc}
            \multicolumn{4}{c}{Decoder} \\
            \hline
            & Layer & \multicolumn{1}{c}{Configuration} & Output shape \\
            \hline\
            
            \multirow{2}{*}{\rotatebox[origin=c]{90}{$S_5$}} & Up & f:(2,2,2) & 24$\times$16$\times$8 \\
                                                             & Conv & k:(3,3,3), \#:128 & 24$\times$16$\times$8 \\
            \hline
            
            \multirow{2}{*}{\rotatebox[origin=c]{90}{$S_4$}} & Up & f:(2,2,2) & 48$\times$32$\times$16 \\
                                                             & Conv & k:(3,3,3), \#:64 & 48$\times$32$\times$16 \\
            \hline
            
            \multirow{2}{*}{\rotatebox[origin=c]{90}{$S_3$}} & Up & f:(2,2,2) & 96$\times$64$\times$32 \\
                                                             & Conv & k:(3,3,3), \#:32 & 96$\times$64$\times$32 \\
            \hline
            
            \multirow{2}{*}{\rotatebox[origin=c]{90}{$S_2$}} & Up & f:(1,2,1) & 96$\times$128$\times$32 \\
                                                             & Conv & k:(3,3,3), \#:16 & 96$\times$128$\times$32 \\
            \hline
            
            \multirow{3}{*}{\rotatebox[origin=c]{90}{$S_1$}} & Up & f:(1,2,1) & 96$\times$256$\times$32 \\
                                                             & Conv & k:(3,3,3), \#:8 & 96$\times$256$\times$32 \\
                                                             & Conv & k:(1,1,1), \#:2 & 96$\times$256$\times$32 \\
            \hline
            & & &\\
        \end{tabular}
    \end{minipage}
\end{table}

Weak SNR, as well as image artifacts in the SELF-OCT scans, are very challenging for automatic segmentation, which leads to a reduction in segmentation accuracy in these image areas. Based on many years of experience, an ophthalmologist can reconstruct and delineate the shape of a retinal surface, even if the image quality is insufficient. To integrate this anatomical prior knowledge into our segmentation algorithm a convolutional denoising autoencoder (CDAE) was used to refine the U-Net prediction.

Autoencoders are able to learn sparse intermediate representations from which the original input data can be reconstructed again. They are therefore used to learn representations of data sets, applying a dimensional reduction similar to principal component analysis. In addition, a denoising autoencoder aims to reconstruct the original data from corrupted inputs. In this way, robust representations can be learned and the learning of the identity function can be prevented.

As depicted in Fig. \ref{fig:scheme} the CDAE receives the binarized retinal prediction of the U-Net (PED (\swatch{green}) is handled as part of the retina \swatch{red}) as input and outputs a refined segmentation. The architecture of the CDAE is shown in Tab. \ref{tab:cdae_architecture}. The CDAE consists of an encoder-decoder architecture that is similar to that of the U-Net except for the skip connections. In contrast to the U-Net bilinear upsampling operations are used for the CDAE instead of upconvolutions to reduce the number of parameters. In the encoder representative features of the input volume are learned on six scale levels while gradually reducing the image resolution. The result is a representation of the input reduced to a few parameters, which is represented by the last layer of the encoder, also called bottleneck. Subsequently, sparse representation in the bottleneck is successively reconstructed in the decoder path. 

The CDAE was trained in the same way as the U-Net. Also the generalized Dice was used as loss function. To ensure that the CDAE learns to reconstruct a denoised version of its input data we performed a series of random augmentations during training. Binary noise, as well as elastic transformations, were applied to the input volume. Furthermore, randomly generated ellipsoids were added or subtracted to the input volume. The refined retinal shape was then fused with the U-Net prediction to generate the final segmentation. If the U-Net prediction does not contain any PED structures, only the refined retina (\swatch{red}) by the CDAE was used as a final segmentation. Otherwise, the multi-label prediction of the U-Net was masked with the refined retinal shape of the CDAE. PED (\swatch{green}) areas are adjusted if the lower retinal boundary has shifted downwards. In this case, the PED area was extended to this new lower retinal boundary.

\section{Experiments and results}
\label{sec:experimentsandresults}
The segmentation results of the U-Net were compared with the refined ones by the CDAE. 5-fold cross-validation was carried out for this purpose. For each run, the image data was divided into training, validation and test data sets in a ratio of 60\%-20\%-20\%. The Dice similarity coefficient (DSC) was used as a metric for quantitative evaluation. In addition, the average symmetric surface distance (ASSD) and the Hausdorff distance (HD) were computed for the retina (\swatch{red}). Note that the image resolution was reduced in width to speed up the training time and to obtain almost equal pixel spacing in both lateral dimensions.

\begin{figure}[t]
    \centering
    \begin{tabular}{cc}
        \includegraphics[width=0.47\textwidth]{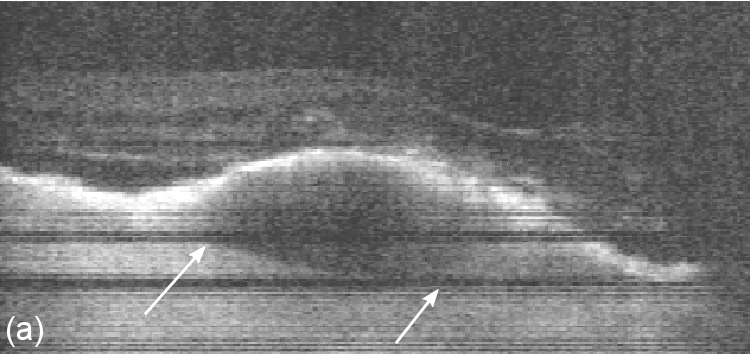} & \includegraphics[width=0.47\textwidth]{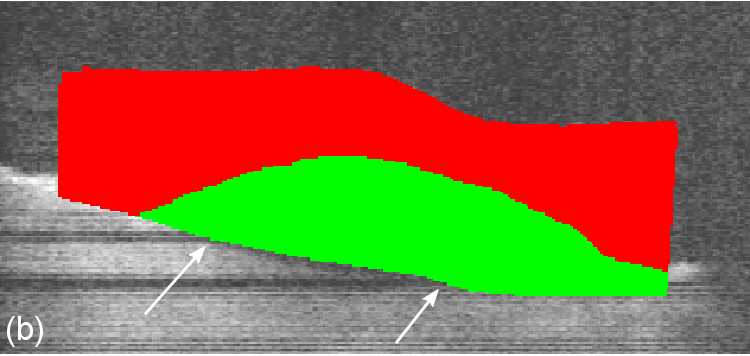} \\
        \includegraphics[width=0.47\textwidth]{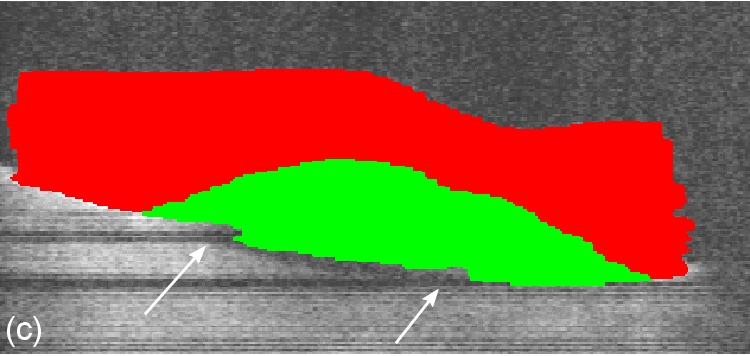} & \includegraphics[width=0.47\textwidth]{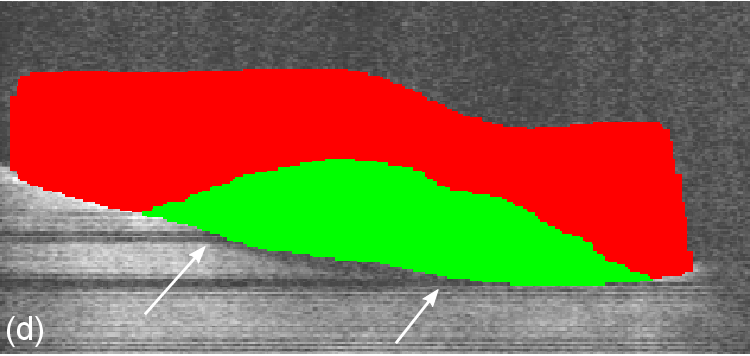} \\
        \includegraphics[width=0.47\textwidth]{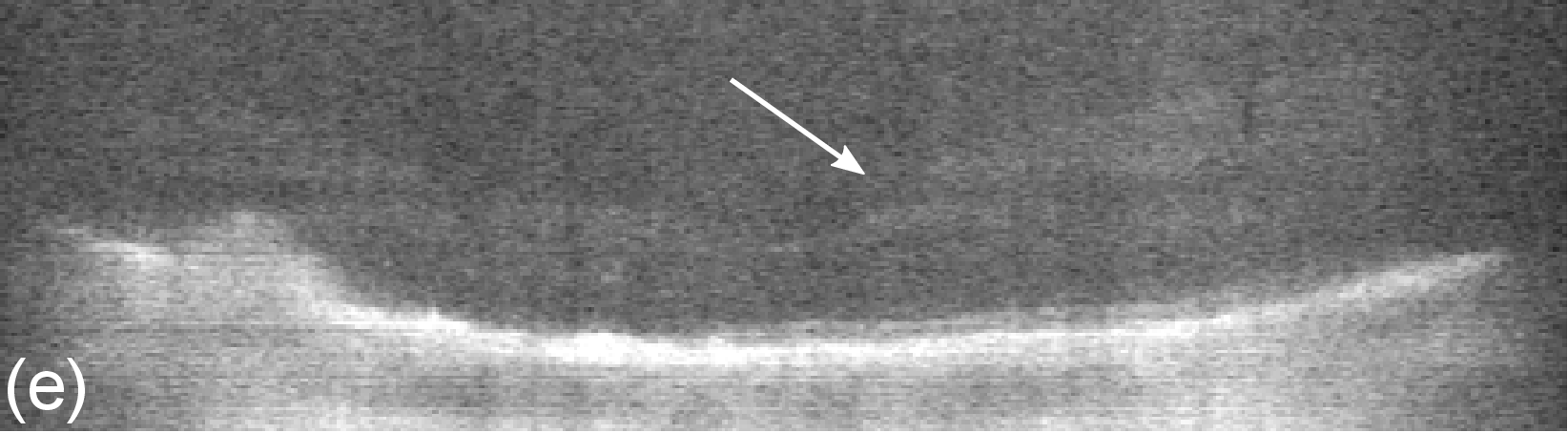} & \includegraphics[width=0.47\textwidth]{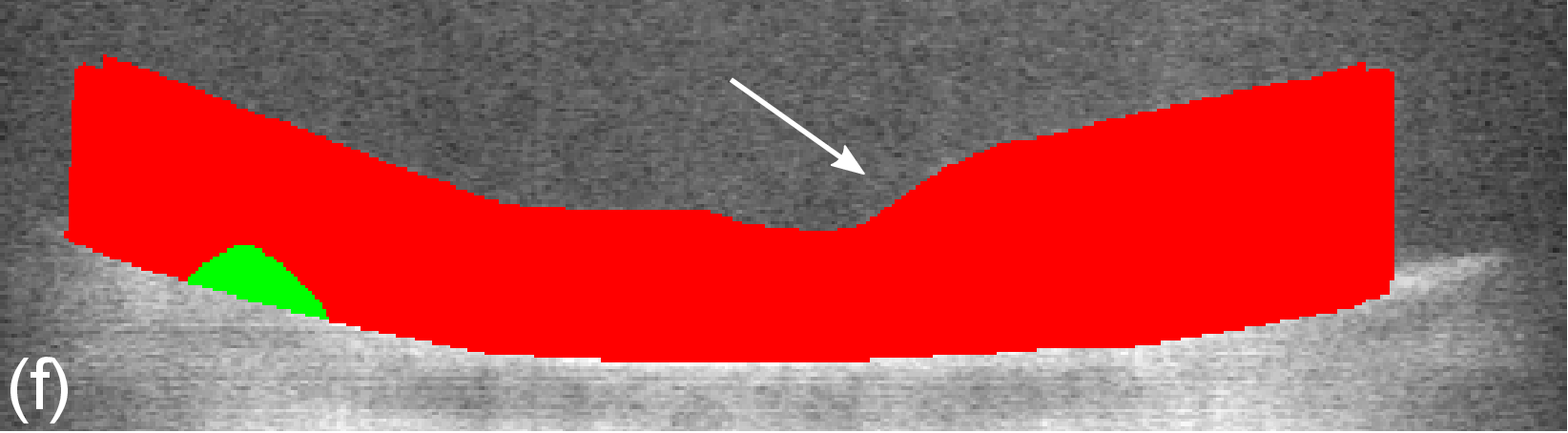} \\
        \includegraphics[width=0.47\textwidth]{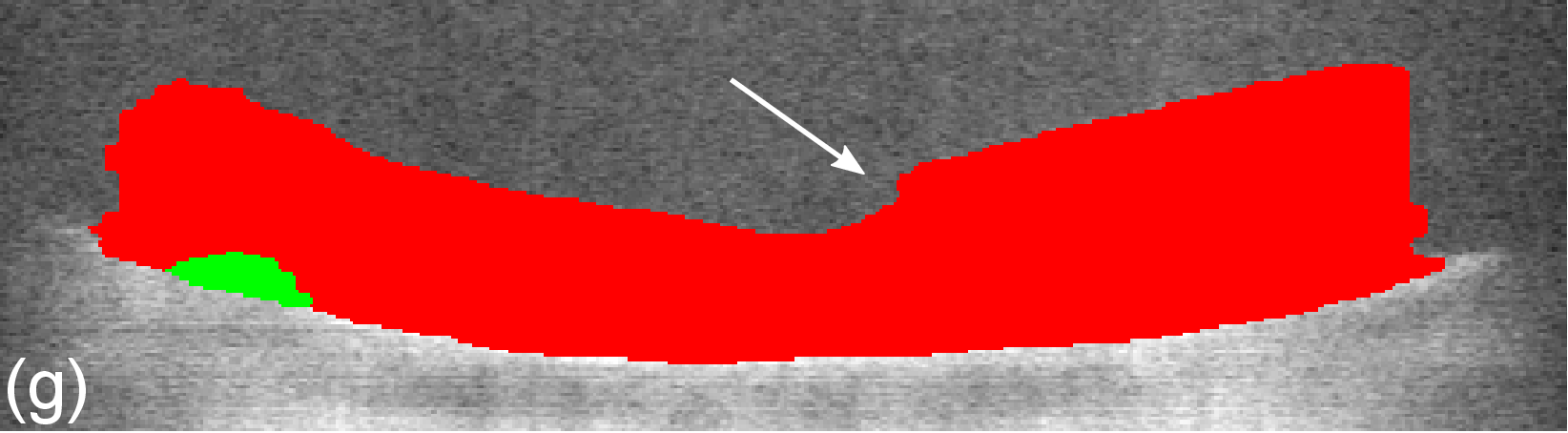} & \includegraphics[width=0.47\textwidth]{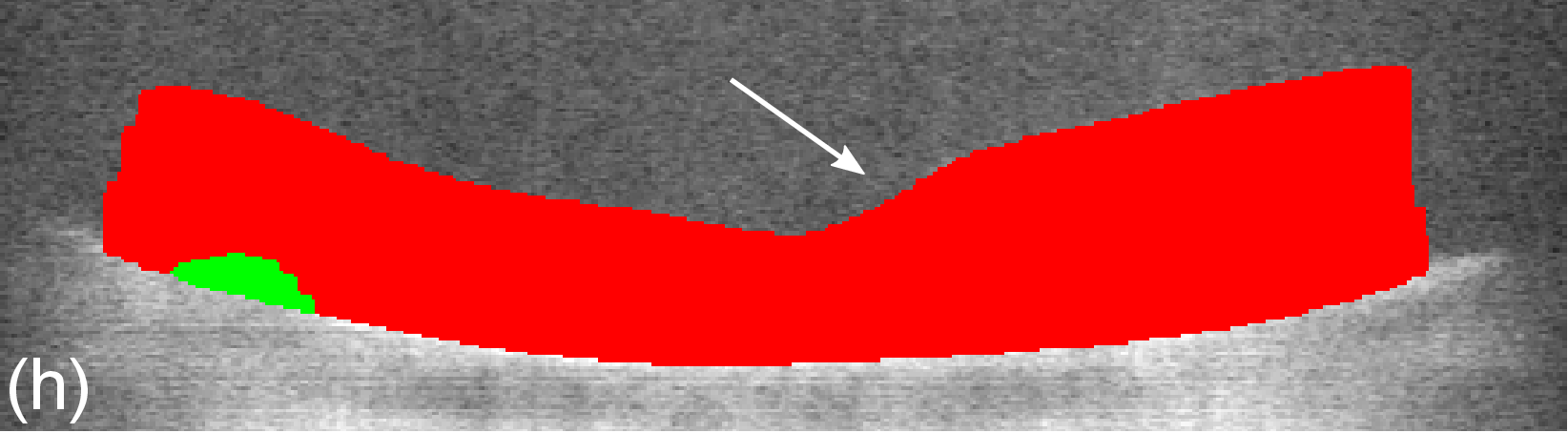} \\
    \end{tabular}
    \vspace{0.5em}
	\caption{Segmentation results of the proposed approach of two retinal B-scans of two individual patients with retina (\swatch{red}) and PED (\swatch{green}): B-scan (a)/(e), ground truth (b)/(f), U-Net segmentation (c)/(g), CDAE segmentation (d)/(h). Areas with low SNR or motion artifacts (white arrow) (a)/(e) can lead to a false segmentation (c)/(g) and are corrected by the CDAE refinement (d)/(h). \label{fig:qualitativeresult}}	
\end{figure}

\begin{table}[h]
    \centering
    \caption{Metric results of the 5-fold cross validation of the segmentation performance. \label{tab:results}}
    \vspace{0.5em}
    \begin{tabular}{c|c|c|c}
         \multicolumn{2}{c|}{Anatomic labels $\rightarrow$}  &  \multirow{2}{*}{Retina (\swatch{red})} & \multirow{2}{*}{PED (\swatch{green})}\\\cline{1-2}
         Metrics $\downarrow$ & Methods $\downarrow$ & & \\\hline
         \multirow{2}{*}{DSC} & U-Net & 0.939 & 0.593 \\
                              & U-Net+CDAE & 0.939 & 0.6 \\\hline
         \multirow{2}{*}{ASSD} & U-Net & 13.26\,$\mu$m & -- \\
                               & U-Net+CDAE & 13.57\,$\mu$m & -- \\\hline
         \multirow{2}{*}{HD}   & U-Net & 274.01\,$\mu$m & -- \\
                               & U-Net+CDAE & 269.61\,$\mu$m & -- \\\hline
    \end{tabular}
\end{table}

Qualitative results are shown in Fig. \ref{fig:qualitativeresult}. Especially axial motion artifacts can cause signal loss, which results in dark horizontal stripe artifacts that negatively affect the segmentation process. It could be shown qualitatively that the CDAE corrects corrupted segmentations. Furthermore, the CDAE generates smoother segmentation surfaces. One can recognize this very well by the outer vertical segmentation boundaries. The metric results of the quantitative evaluation are shown in Tab. \ref{tab:results}. One can notice that the U-Net segments the retina (\swatch{red}) on average with high accuracy (DSC: 0.939, ASSD: 13.26\,$\mu$m, HD: 274.01\,$\mu$m). In contrast, segmentation of PED (\swatch{green}) was more challenging (DSC: 0.593). Compared to the qualitative evaluation, the results of the CDAE refinement show only minimal deviations for retina (\swatch{red}) (DSC: 0.939, ASSD: 13.75\,$\mu$m, HD: 269.61\,$\mu$m) and PED (\swatch{green}) (DSC: 0.6).  

\section{Discussion and conclusion}
\label{sec:discussionandconclusion}
In this paper, a deep learning-based approach for the automatic segmentation of SELF-OCT image data was presented. A 3D-U-Net was used to segment image volumes of the SELF-OCT system. In order to correct possible segmentation errors caused by image artifacts, a CDAE was additionally utilized. For this purpose, the CDAE was trained with noisy binarized ground truth segmentations enabling it to learn to perform a shape refinement. For evaluation, 5-fold cross-validation was conducted. The quantitative evaluation showed that the retina was segmented with high accuracy. Segmentation with respect to the PED was more challenging due to its less clear demarcation in the SELF-OCT images. One reason here for this could be the fact that the BM, which marks the lower boundary of the PEDs, was difficult to detect due to the highly reflective RPE layer located above. Moreover, it was shown qualitatively that the CDAE refinement reduces segmentation errors caused by artifacts using learned prior knowledge. However, the quantitative results of the CDAE refinement showed only minor changes. Since the technology of the SELF-OCT is continuously being optimized, an improvement in image quality can be expected in the future, which will have a positive effect on segmentation performance.


\bibliography{report} 
\bibliographystyle{spiebib} 

\end{document}